\documentclass[usenatbib]{mn2e}
\usepackage[letterpaper,margin=0.8in]{geometry}
\usepackage{hyperref}
\usepackage{natbibmnfix,fixltx2e}
\usepackage{astrojournals}
\usepackage{amsmath,amssymb,mathtools}
\usepackage{microtype}
\usepackage{txfonts}
\usepackage{graphicx}
\usepackage{xspace}

\renewcommand*{\vec}[1]{\ensuremath{\boldsymbol{#1}}}
\let\oldhat\hat
\renewcommand*{\hat}[1]{\oldhat{\vec{#1}}}

\renewcommand{\c}[2]{\ensuremath{#1_{\text{#2}}}} 
\newcommand{\ia}{\c{i}{a}\xspace}
\newcommand{\ib}{\c{i}{b}\xspace}
\newcommand{\ie}{\c{i}{e}\xspace}

\newcommand{\unitvectorslope}[1]{\ensuremath{%
\frac{\c{\oldhat{#1}}{z}}%
     {\sqrt{1-\c{\oldhat{#1}}{z}^2}}}}

\def\deg{\hbox{$^\circ$}}

\hypersetup{%
  pdftitle={A new inclination instability in Keplerian disks},
  pdfauthor={\textcopyright\ authors},
  bookmarksopen=true,
  colorlinks=true,
  linkcolor=black,
  citecolor=black,
  urlcolor=black}

\begin{document}

\title[Inclination Instability]{A new inclination instability reshapes Keplerian disks into cones: application to the outer Solar System}

\author[Madigan \& McCourt]
{Ann-Marie Madigan$^1$\thanks{ann-marie@astro.berkeley.edu} and Michael McCourt$^2$ 
\vspace{0.1in} \\
$^1$ Astronomy Department and Theoretical Astrophysics Center, University of California, Berkeley, CA 94720, USA \\
$^2$ Institute for Theory and Computation, Harvard University, Center for Astrophysics, 60 Garden St., Cambridge, MA 02138 }
\maketitle

\begin{abstract}

Disks of bodies orbiting a much more massive central object are extremely common in astrophysics. When the orbits comprising such disks are eccentric, we show they are susceptible to a new dynamical instability.  Gravitational forces between bodies in the disk
drive exponential growth of their orbital inclinations and clustering
in their angles of pericenter, expanding an initially thin disk into a
conical shape by giving each orbit an identical ``tilt'' with respect
to the disk plane.  This new instability dynamically produces the
unusual distribution of orbits observed for minor planets beyond
Neptune, suggesting that the instability has shaped the outer Solar
System. It also implies a large initial disk mass ($\sim\,1$\,--\,$10$ Earth masses) of scattered bodies at hundreds of AU; we predict increasing numbers of detections of minor
planets clustered in their angles of pericenter with high
inclinations.

\end{abstract}

\begin{keywords}
celestial mechanics --- planets and satellites: dynamical evolution and stability ---  minor planets --- stars: kinematics and dynamics
\end{keywords}

\section{Introduction}
\label{sec:intro}
Disks of bodies orbiting a more massive central object are ubiquitous
in the universe: examples include moons around a planet, planets
around a star, or stars around a super-massive black hole.  The
long-lived, periodic interactions among closed orbits in these disks
can drive rapid angular momentum evolution among bodies comprising the disk (e.g., ``resonant-relaxation'' of stars near massive black holes;  
\citealt{Rauch1996}) and collective behavior such as spontaneous organization or dynamical instability \citep{Madigan2009}.  While particles in circular disks do not
undergo collective behavior and are understood to diffuse slowly in
inclination only through two-body scattering \citep{Ida1990}, we show
here that disks of particles with high eccentricity
orbits instead collect in angle of pericenter and undergo exponential growth in inclination; we refer
to this new process as the \textit{inclination instability}.

In this letter, we demonstrate the inclination instability and illustrate its essential characteristics; we also present evidence that the
  outer Solar System has undergone the inclination instability, and
  that it reproduces the bizarre and unexplained
  distribution of orbits of minor planets beyond Neptune.
We describe the secular dynamical mechanism by which the instability works in a forthcoming paper (Madigan \& McCourt, \textit{in prep.}).  

\section{The Inclination Instability}
\label{sec:instability}

In order to demonstrate the inclination instability, we consider a
thin, axisymmetric disk of particles on eccentric orbits around a much
more massive central object.  Two angles are required to orient an
orbital plane in space;\footnote{Traditionally the longitude of
  ascending node $\Omega$ and inclination angle $i$.} we adopt two
inclination angles \ia and \ib.  These angles represent rotations of
the orbit with normal vector $\hat{j}$ about its semi-major ($\hat{a}$) axis and its semi-minor
($\hat{b} \equiv \hat{j}\times\hat{a}$) axis, respectively:
\begin{subequations}\label{eq:am-inc-definition}
\begin{align}
  \tan(\ia) &=  \unitvectorslope{b}\label{eq:ia-def} \\
  \tan(\ib) &= -\unitvectorslope{a}\label{eq:ib-def}
\end{align}
\end{subequations}
Though non-standard, the angles \ia and \ib are characteristic
coordinates of the inclination instability and greatly simplify our
analysis.

\begin{figure*}
  \includegraphics[width=\linewidth]{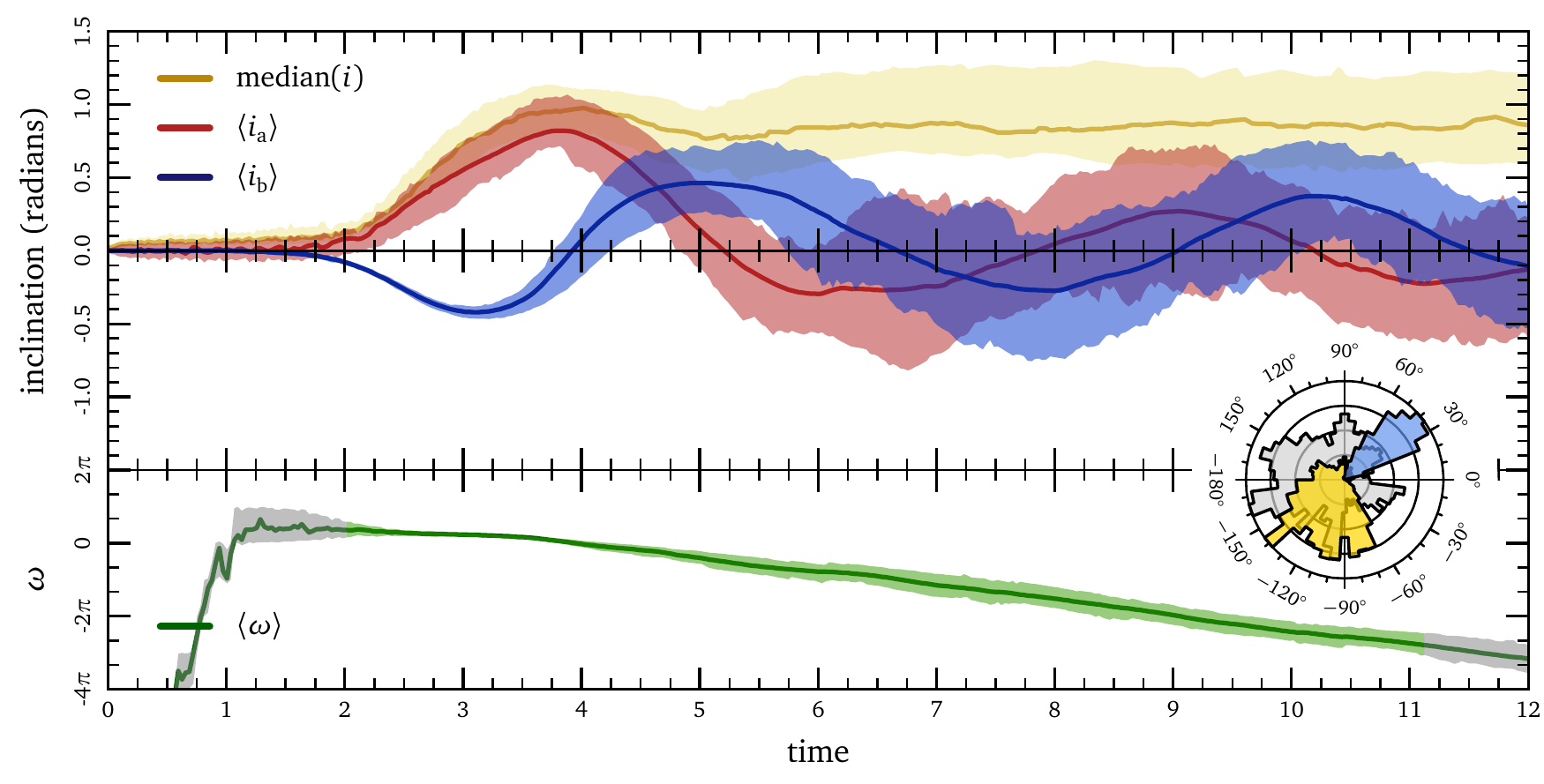}
  \caption{(\textit{top}): Evolution of the inclination angles \ia and
    \ib (defined in equation~\ref{eq:am-inc-definition}) for particles
    initially forming a thin, near-Keplerian disk.  Lines show
    median/mean angles and the shaded region encloses the 25th and
    75th quantiles, i.e. enclosing 50\% of the particles.  Orbital
    inclinations initially grow rapidly due to the inclination
    instability.  Each orbit has approximately the same inclination,
    with \ia and \ib initially growing together in a fixed ratio $<0$.
    After the instability saturates around a time $t\sim3$, the orbits
    precess coherently about the mean angular momentum axis of the
    disk.  (\textit{bottom}): Evolution of the angle of pericenter
    $\omega$.  The inclination instability causes the orbits to
    collapse into a narrow distribution of $\omega$ around a time
    $t\sim2$\,--\,$3.$ This clustering in $\omega$ is not actively
    maintained after the instability saturates, and spreads due to
    differential precession by a time $t\sim6.5$.  As in the top
    panel, the colored region shows the width of the distribution
    enclosing half of the orbits.  We mark this as green where the
    width is less than $\pm55\deg$, the observed scatter for minor
    planets beyond Neptune (see section~\ref{sec:discussion}, below). (\textit{inset}):
    angular distribution of $\omega$ during the linear, exponentially
    growing phase from $t\sim2.5$\,--\,$3$ (blue) and in the saturated
    state ($t\sim5.5$\,--\,$6$; yellow).  The asymmetry in $\omega$
    persists even at late times ($t\sim11.5$\,--\,$12$;
    gray).}\label{fig:iaib}
\end{figure*}

\begin{figure}
  \includegraphics[width=\linewidth]{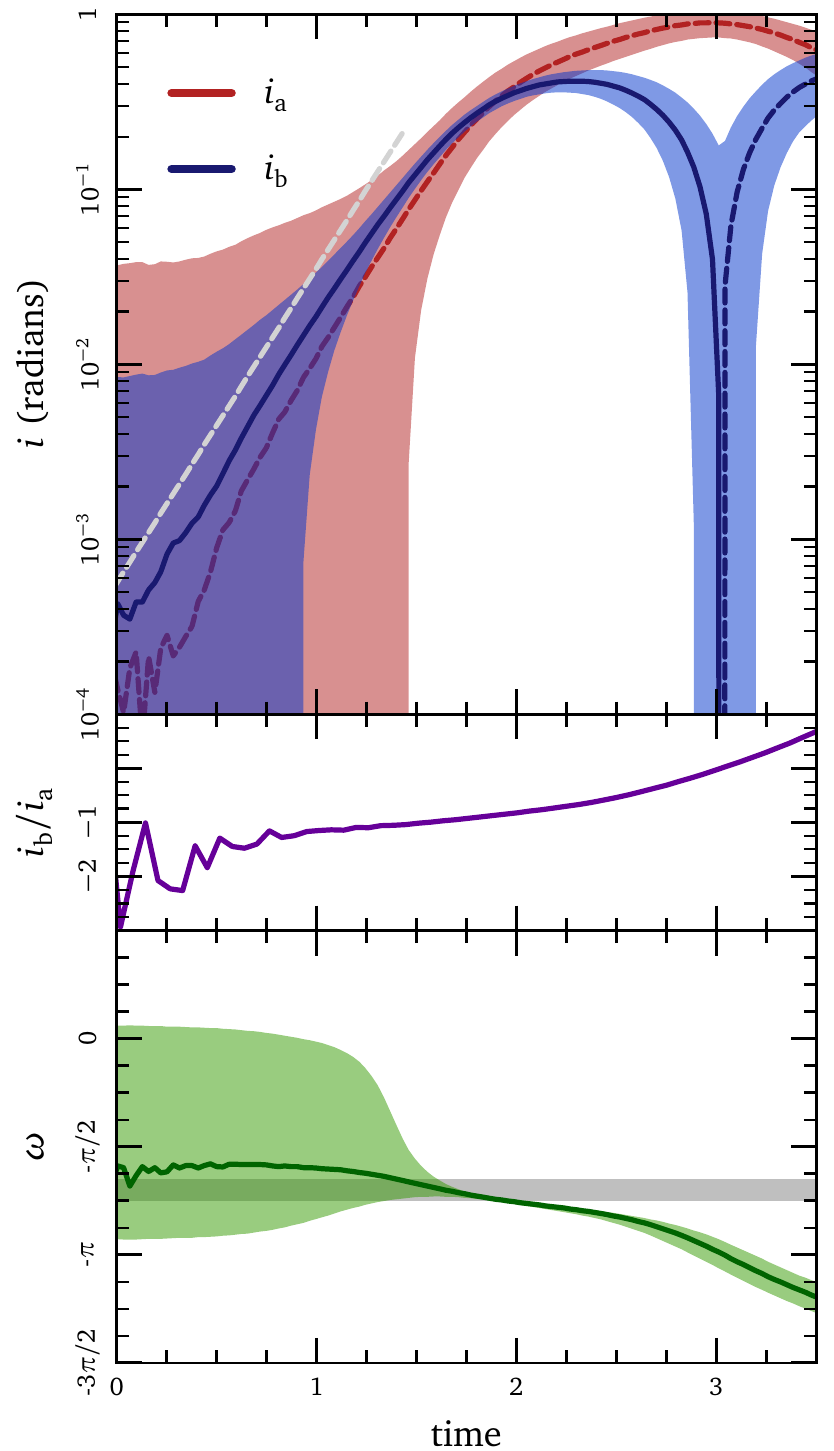}
  \caption{Quantitative evolution of the inclination
    instability. (\textit{top}): Exponential growth of the inclination
    angles $\ia$ and $\ib$.  Lines show the median values of all the
    orbits and the colored bands show the width of the distribution
    enclosing 50\% of the orbits.  Solid lines indicate positive
    values, and dashed lines indicate negative values.  At early
    times, the individual angular momentum vectors precess
    incoherently; this precession implies oscillations in \ia and \ib
    with $|\ia| > |\ib|$.  The precession averages to zero upon taking
    the median, revealing exponential growth with $|\ib| > |\ia|$.
    The instability is present even at low inclinations ($< 1\deg$)
    though this may not be observable in systems of small numbers of
    bodies. (\textit{middle}): Evolution of the ratio
    $\langle\ib\rangle/\langle\ia\rangle$.  Though $\langle\ib\rangle$
    and $\langle\ia\rangle$ individually grow by more than two orders
    of magnitude from $t\sim0$\,--\,$1.5$, their ratio remains
    constant to within a factor of 2.  (\textit{bottom}): Evolution in
    mean angle of pericenter.  We over-plot a theoretical prediction.
    The colored bands show the width of the distribution enclosing
    50\% of the orbits.  The distribution in $\omega$ remains tight
    even after the exponential growth phase ends. }
\label{fig:linear-phase}
\end{figure}

Figure~\ref{fig:iaib} shows the evolution of the inclination angles
\ia and \ib in an $N$-body simulation of low-mass bodies orbiting a
central object of mass $M$.  We initialize $N=100$ particles
in an axisymmetric disk with a total mass $M_{\text{disk}} = 10^{-4}
M$ and identical eccentricities $e$=0.9. 
We perform integrations using the \textit{mercury6} code \citep{Cha99}.
We normalize time to the secular dynamical timescale
$t_{\text{sec}} \equiv (M/M_{\text{disk}})\,P$, where $P$ is a typical
orbital period.  The top panel
of figure~\ref{fig:iaib} shows that the inclination angles initially
grow rapidly, with opposite signs $\ib/\ia < 0$ and with $|\ia| <
|\ib|$.  All three of these features are inconsistent with known
processes such as scattering or resonant relaxation;
to our knowledge, this behavior is uniquely characteristic of the
inclination instability.

The instability saturates, and exponential growth of orbital
inclination halts, when inclinations reach $\sim{}1$\,radian.  After this time, we see large-scale,
coherent precession of the orbits' individual angular momentum vectors
about the mean angular momentum of the disk; this large-scale
precession produces the slow oscillations in \ia and \ib seen at late
times in figure~\ref{fig:iaib}, while preserving the overall
inclinations of the orbits [$i\equiv\arccos(j_{z}/j)$].

A direct observational signature of the inclination instability is the
distribution of orbits in angle of pericenter\footnote{$\omega =
  \arccos{{\vec{\hat n}}.{\vec{\hat e}}}$, is the angle between the
  vector of ascending node and eccentricity vector.}  $\omega$: The
bottom panel of figure~\ref{fig:iaib} shows that, during the
exponentially growing phase of the instability, this distribution of
$\omega$ values collapses to an extremely narrow range
$\omega\sim(45\pm15)\deg$.  The distribution remains narrow even after
the exponential phase of the instability ends, though it eventually
spreads out due to differential axial precession of orbits within the disk.
Physically, a tight distribution in $\omega$ implies that orbits tilt
uniformly out of the disk plane, giving the disk a conical shape; see
figure~\ref{fig:omega-diagram} for a 3D visualization.  The top panel
of figure~\ref{fig:iaib} shows that the inclination instability
creates this shape by simultaneously giving each orbit similar values
for \ia and \ib.  This tilt moves the center of mass of the disk off
the original disk plane and drives the late-time precession seen in
the top panel of figure~\ref{fig:iaib}.

The bottom panel of figure~\ref{fig:iaib} shows that the inclination
instability drives $\omega$ to a constant value $\sim{}45\deg$ during
its linear, exponentially-growing phase from $t\sim1$ until
$t\sim3\text{ish}$.  Before this time, orbits individually undergo
retrograde axial precession.\footnote{while the orbits also undergo
  apsidal precession, the rate of this precession is much lower for
  the high eccentricities considered here.} This axial precession is
driven by self-gravity of the disk, which introduces a non-keplerian
force on particles toward the disk plane; consequently, $\omega$
increases roughly linearly with time before the instability sets in.  
After the inclination instability collects orbits in angle of
pericenter (effectively rearranging the disk into a conical shape), the dominant non-keplerian
force no longer points toward the disk plane, but instead toward the
disk's axis of symmetry, above or below the disk plane.  This force
drives \textit{prograde} axial precession, reversing the trend in
$\omega$ as seen in figure~\ref{fig:iaib} after $t\sim3\text{ish}$.

Figure~\ref{fig:linear-phase} further quantifies the development of
the inclination instability.  The top panel shows the inclination
angles \ia and \ib as functions of time during the linear phase of the
instability, clearly demonstrating the exponential growth expected for
a true dynamical instability.  At early times in the simulation, the
angular momentum vectors precess rapidly and incoherently about the
mean angular momentum of the disk; this precession represents
oscillations in \ia and \ib with $|\ia| > |\ib|$, and it dominates the
orbital inclinations until the inclination instability takes over
around a time $t\sim1$\,--\,$1.5$.  The colored bands in the top panel
of figure~\ref{fig:linear-phase} indicate the width of the
distribution by enclosing 50\% of the orbits; this indicates the
amplitude of the precession.  The initial precession is incoherent,
however, and its effects average away upon taking
the median.  The thick curves in the top panel of
figure~\ref{fig:linear-phase} thus reveal exponential growth with
$|\ib| > |\ia|$ even at early times $t<1$, when the instability is
much lower in amplitude than the precession.  (We include the results
of 1024 independent simulations to reduce the noise in this average).

The physical mechanism behind the inclination instability implies a
constant (but eccentricity-dependent) ratio $\ib/\ia \lesssim -1$
(Madigan \& McCourt, \textit{in prep.}).  The two curves in the top panel
of figure~\ref{fig:linear-phase} show \ia and \ib growing at a
constant, negative ratio at early times.  We demonstrate this
explicitly in the middle panel, where we plot the ratio
$\langle\ib\rangle/\langle\ia\rangle$; though the curve is noisy due
to precession (larger in magnitude by a factor of $\sim100$), this
ratio is consistent with the theoretical model.  Though
$\langle\ib\rangle$ and $\langle\ia\rangle$ individually grow by more
than two orders of magnitude during the linear, exponentially-growing
phase of the instability, their ratio remains constant to within a
factor of 2.

For small inclinations, a constant ratio $\ib/\ia$ implies a constant
angle of pericenter $\omega\sim\arctan(|\ib/\ia|)$ ($+\pi$ if
$\ia<0$).  The bottom panel of figure~\ref{fig:linear-phase} shows the
angle of pericenter $\omega$ as a function of time in our simulation,
along with our theoretical expectation (Madigan \& McCourt \textit{in prep.}).  As in the top panel, the
colored region encloses 50\% of the orbits and quantifies the scatter
in the simulation.  Individual orbital precession dominates the
inclinations until a time $t\sim1$ and leads to a large scatter of
$\sim{}2\pi$; however, the distribution tightens considerably once the
inclination instability dominates the dynamics.  As seen in
figure~\ref{fig:iaib}, the distribution remains narrow even after the
exponential phase of the instability ends around the time $t\sim2$.

Figure~\ref{fig:omega-diagram} shows the geometric meaning of this
narrow distribution in $\omega$.  The middle and bottom panels show
two disks in which every orbit has the same semi-major axis $a$,
eccentricity $e=0.8$, and inclination $i=0.5$.  In the middle panel,
every orbit has the same value of $\omega=-1$; the orbits in this disk
coherently tilt above the disk plane to form a cone shape.  (For
positive values of $\omega$, the orbits would tilt below the plane.)  
In the bottom panel, we give each orbit a random value of
$\omega$; this disk thickens isotropically into a torus.

\begin{figure*}
  \includegraphics[width=\linewidth]{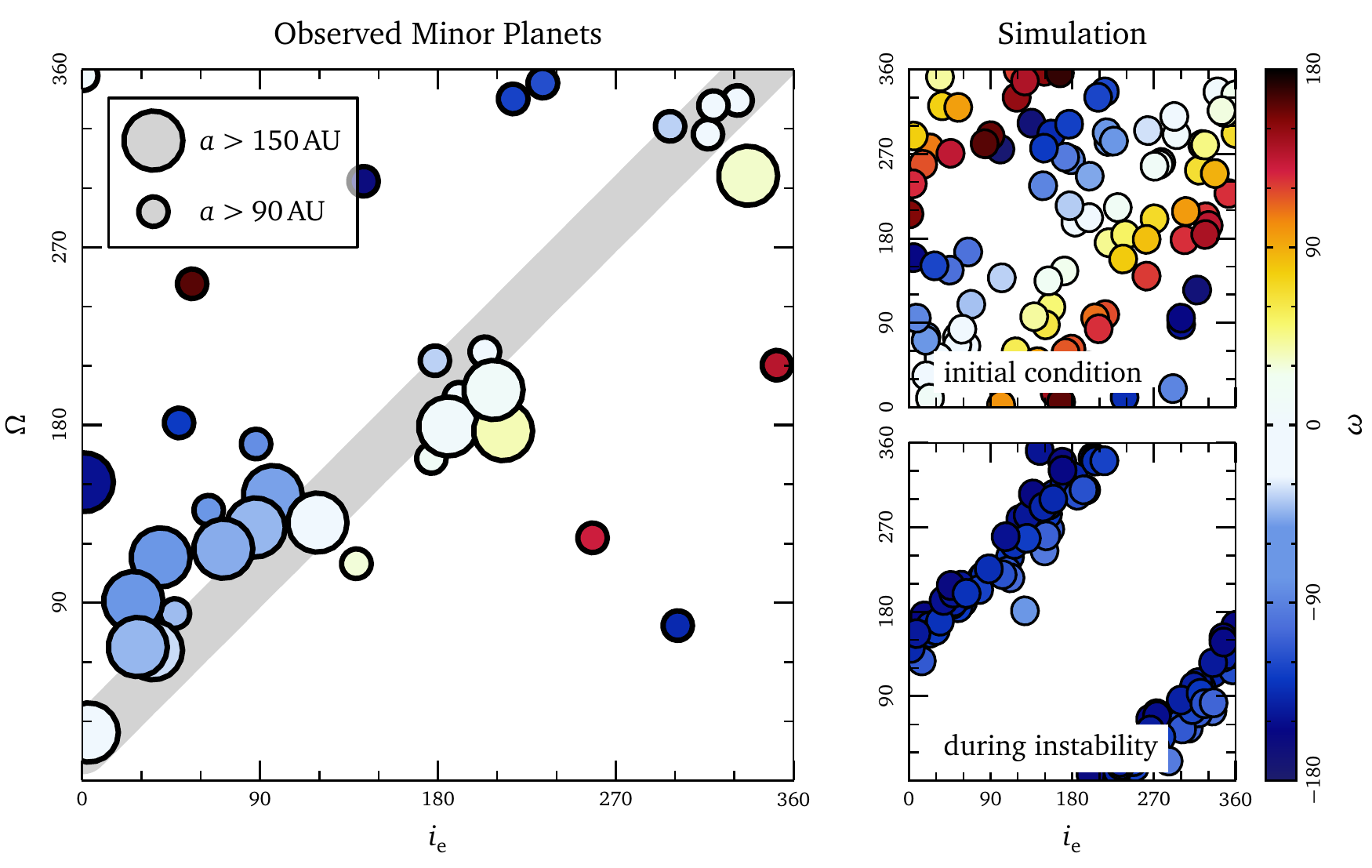}   
      \caption{Inclination instability in the outer Solar System.
        (\textit{left}): orientation angles measured for minor planets
        in the solar system.  We show all known minor planets with
        pericenter distances greater than Neptune's orbit (30\,AU) and
        with semi-major axes greater than 90\,AU (small circles) and
        150\,AU (large circles, cf.~\citet{Trujillo2014}).  We plot
        the longitude of ascending node $\Omega$ and the angle $\ie
        \equiv \arctan( \c{a}{y}, {a_x})$, which represents the
        orientation of the semi-major axis (or eccentricity vector)
        within the reference plane.  Orbits randomly distributed
        within a disk would fill out this space, however the minor
        planes only occupy a narrow band.  Error bars are smaller than
        symbols.  (\textit{right}): \textit{N}-body simulation of the
        inclination instability in a disk with 100 particles and a
        total mass $M_{\text{disk}} = 10^{-4}\,M$.  The top panel
        shows the initial condition for the simulation; orbits
        randomly distributed within the disk populate the entire
        $\Omega$\,--\,$\ie$ space.  During the inclination
        instability, this distribution collapses to a narrow band in
        $\Omega$\,--\,$\ie$ space, as in seen in the outer solar
        system.  In all plots, color indicates the angle of pericenter
        $\omega$; banding in $\Omega$\,--\,$\ie$ space corresponds to
        clustering in $\omega$.}
    \label{fig:dataCompSims}
\end{figure*}

\section{Discussion}
\label{sec:discussion}
\begin{figure}
  \centering
  \includegraphics[width=\linewidth]{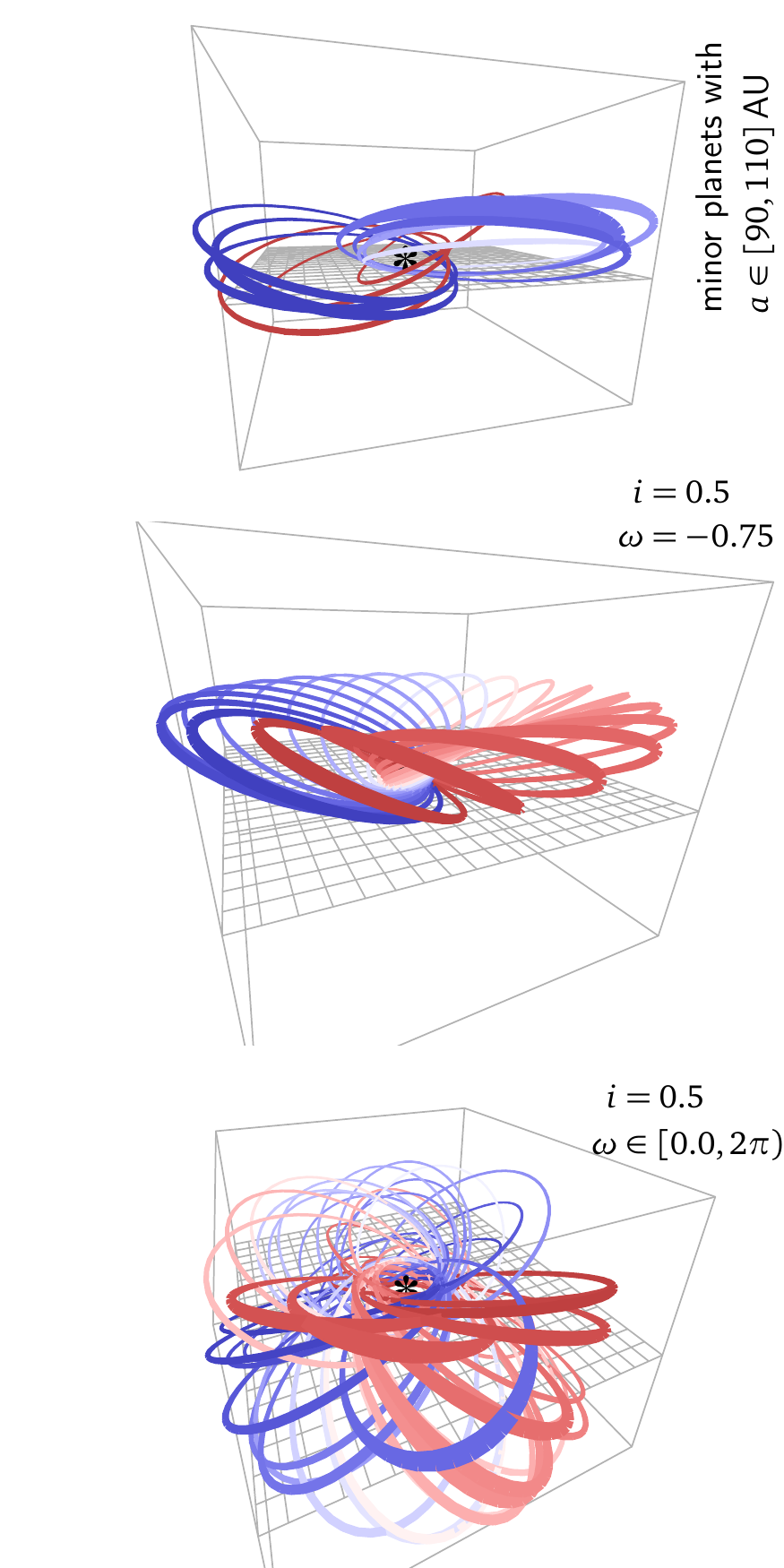}
  \caption{Geometric meaning of ``clustering in $\omega$.''
    (\textit{middle}): a disk of 20 orbits, all with identical
    semi-major axis $a$, eccentricity $e=0.8$, inclination $i=0.5$,
    and angle of pericenter $\omega=-0.75$.  Longitude of ascending node
    $\Omega$ uniformly fills $[0,2\pi)$ and is indicated with color.
      Disks which cluster in $\omega$ represent ``cones'' with the
      center of mass above or below the disk plane, and with each
      orbit tilting in the same fashion.  (\textit{bottom}): the same
      disk, but each orbit is given a random value of
      $\omega\in[0,2\pi)$; in this (much more generic) case, the disk
        becomes a thick torus.  (\textit{top}): Orbits for minor
        planets with semi-major axis lengths between 90 and 110\,AU;
        the disk clusters in $\omega$ and has a cone-shape as seen in
        the middle panel.  (We restrict the semi-major axis lengths to
        a narrow range for clarity; the results look similar for other
        ranges.)}\label{fig:omega-diagram}
\end{figure}
The tight distribution in $\omega$ (or corresponding conical shape to
the disk) is a unique observational signature of the inclination
instability.  \citet{Trujillo2014} report such a signature in
observations of minor planets beyond Neptune ($a = 30$ AU).  These icy
bodies are hypothesized to have formed much closer to the sun,
scattering off Neptune to distant, high eccentricity orbits during the
migration phase of the gas- and ice-giants \citep{Tsiganis2005}.
Trujillo and Sheppard show that the angles of pericenter for minor
planets with semi-major axes $a > 150$\,AU and pericenter distances
$r_{\text{peri}} > 30$\,AU cluster about $\langle\omega\rangle=
(340\,\pm\,55)\deg$ (see the top panel of
figure~\ref{fig:omega-diagram}).  This clustering cannot be explained
by any known observational biases.  The authors suggest that a
stochastic event, such as a strong stellar encounter early in the
Solar System's evolution, could produce an initially asymmetric
population in $\omega$.  To prevent $\omega$ values from subsequently
randomizing, they invoke a massive unseen perturber, such as a
trans-Plutonian super-Earth at $a\sim250$\,AU.

We propose a simpler solution: the 
minor planets underwent the inclination instability $\sim$1\,--\,4 Gyr
ago, exponentially growing in inclination, and developing a narrow
distribution in $\omega$ centered on $45\deg$ (for very eccentric
orbits) to $\sim70 \deg$ (for lower eccentricity).  Afterwards, the
distribution of $\omega$ values drifted clockwise and widened due to
precession, before reaching the current value of
$\langle\omega\rangle= (-20\,\pm\,55)\deg$.\footnote{Depending on the
  precession rate and the time since the instability, both of
  which depend on the unknown mass of the disk, $\langle\omega\rangle$
  could equally have begun between $-45\deg$ and $0\deg$, drifting
  through $\sim2\pi$ to reach its current state.}

Clustering in $\omega$ implies a tight correlation between the angle
of ascending node $\Omega$ and \ie, the orientation of the semi-major
axis (or eccentricity vector) as projected in the initial disk plane;
i.\,e., $\ie \equiv \arctan({a_y}, {a_x})$.  We show this in
figure~\ref{fig:dataCompSims}, where we compare observational data
with results from our $N$-body simulations.  In the left panel, we
plot the observed $\Omega$\,--\,\ie distribution for minor planets,
with the gray line corresponding to a constant $\omega=340\deg$.  Data
is taken from the IAU Minor Planet Center.\footnote{\href{http://www.minorplanetcenter.net}{http://www.minorplanetcenter.net}}  We select minor planets
with $r_{\text{peri}}>30$\,AU and $a>150$\,AU, as in
\citet{Trujillo2014}.  In order to show that this clustering is not an
artifact of any particular cut to the data, we also plot all minor
planets with $a>90$\,AU.  Orbits randomly placed within a disk would
populate this entire plot; however the minor planets show a clear
correlation between $\Omega$ and \ie.

In the right panels of figure~\ref{fig:dataCompSims}, we plot the
$\Omega$\,--\,\ie distribution for an $N$-body simulation in the
initial condition (\textit{top}) and during the inclination
instability (at time $t\sim1$; \textit{bottom}).  At early times,
bodies in the disk fill out the entire $\Omega$\,--\,\ie space.
During the instability, however, the distribution collapses to a thin
band, with a shape and thickness that resembles the minor planet data.  After the exponential phase ends, the band sweeps slowly across
the plot as $\omega$ values drift due to large-scale precession.  The
green shaded region in the bottom panel of figure~\ref{fig:iaib} shows
the time during which the $\omega$ values lie within the scatter
observed for minor planets in our solar system; the distribution in
$\omega$ remains tight for another $\sim10$ secular timescales after
the instability saturates.  This length of time is likely to be far
longer than the age of the Solar System (however, this timescale may be shortened by perturbations from the gas giants or from the galactic tidal field, not included here).

In addition to explaining the clustering of minor planets in argument
of pericenter, the inclination instability furthermore reproduces the
high orbital inclinations of minor planets with respect to the
ecliptic [$i~\sim~(5-30)\deg$]. Even more suggestive is the paucity of planets with low inclinations; this cannot be understood as the result of scattering, but is a natural outcome of the inclination instability.  The top panel of
figure~\ref{fig:iaib} shows that inclinations remain high for long
periods of time, possibly indefinitely, after being pumped by the
instability.

The inclination instability also provides a natural explanation for
``detached'' minor planets, which have pericenter distances larger (by
$\gtrsim25\%$) than Neptune's orbit \citep{Gladman2002,Iorio2007}.
Since the effectiveness of gravitational scattering from the gas and
ice giants strongly decreases for pericenter distances above $35$\,AU
\citep{Duncan1987}, detached objects cannot be explained through
scattering.  However, detachment is a natural outcome of the
inclination instability: as orbital inclinations grow, eccentricities
drop due to conservation of angular momentum, and pericenter distances
[$\propto(1-e)$] thus increase beyond the reach of the giant planets
in the inner Solar System.  We study the detailed interaction between
scattering and the inclination instability early in this detachment
process in an upcoming paper (Madigan, \textit{in prep.}).

The inclination instability naturally explains the clustering in
$\omega$ of the minor planets beyond Neptune, and the corresponding
correlation in $\Omega$\,--\,\ie space, the high orbital inclinations
of these objects, and the population of detached minor planets.  Thus,
\textit{the inclination instability qualitatively reproduces the
  distribution of minor planet orbits, beginning from a generic
  initial condition.}  This scenario implies an initial total disk mass of
$\sim\,1$\,--\,$10$ Earth masses distributed between $\sim50$\,AU and
$\sim10^4$\,AU, at least an order of magnitude larger than the
estimated current mass in the Kuiper Belt at smaller radii
($a\sim30$\,--\,$50$\,AU) \citep{Gladman2001}.

Existing simulations of the formation of the outer Solar System
predict very few minor planets at distances between $\sim50$\,AU and
$\sim10^{4}$\,AU \citep{Duncan1987}.  However, there are few
observational constraints on this region, as it is an empty loss-cone
regime \citep{Hills81}.  Moreover, in recent years, nearly two hundred
minor planets have been discovered with semi-major axes of hundreds of
AU, despite the extreme difficulty of finding such objects.  It thus
seems entirely possible that earlier theoretical predictions are
incorrect, and that a massive reservoir of minor planets exists beyond
the Kuiper Belt.  We predict increasing numbers of detections of minor
planets clustered in $\omega\sim(-20\pm55)\deg$ values, with high
inclinations and semi-major axes of $\sim100$s of AU.  A further
prediction of the inclination instability is that
$d\langle\omega\rangle/da > 0$; that is, mean $\omega$ values of minor
planets should increase with semi-major axis.  In an upcoming paper, we make more specific, statistical
predictions for where new minor planets should be found.

We close by noting that the inclination instability also has
implications for the dynamics of stars orbiting super-massive black
holes in galactic nuclei.  For example, the disk of young stars in our
Galactic center is much thicker than can be understood via scattering
or resonant relaxation \citep{Yelda2014}, suggesting that the
inclination instability may have expanded the disk.  We explore this
possibility in future work.

\vspace{0.3cm}
\textbf{Acknowledgements} A.-M.M. thanks C. Nixon, M. Jalali,
A. Parker, B. Sherwin and F. Van de Voort for helpful conversations.
We thank R. Murray-Clay, E. Chiang and E. Quataert for
comments on an earlier draft.  A.-M.M. was supported by NASA through
Einstein Postdoctoral Fellowship Award Number PF2-130095.  M.M. was
supported by the National Science Foundation (NSF) grant AST-1312651 and NASA grant NNX15AK81G.  The
authors acknowledge the Texas Advanced Computing Center (TACC) for providing HPC resources that have
contributed to the research results reported here.  This
work used the Extreme Science and Engineering Discovery Environment
(XSEDE allocations TG-AST140039, TG-AST140047, and TG-AST140083),
which is supported by NSF grant number
ACI-1053575.  We used the open-source software
\href{http://tioga.sourceforge.net/}{tioga} to make our plots.

\end{document}